\documentstyle[12pt,amssymb,latexsym,amsmath]{article}
\newcommand{\beq}{\begin{equation}}
\newcommand{\eeq}{\end{equation}}
\newcommand{\ba}{\begin{array}}
\newcommand{\ea}{\end{array}}
\begin{document}
\title{Partially integrable systems in multi-dimensions\\
by a variant of the dressing method. 1}

\author{
A.I. Zenchuk\\
Center of Nonlinear Studies of L.D.Landau Institute
for Theoretical Physics  \\
(International Institute of Nonlinear Science)\\
Kosygina 2, Moscow, Russia 119334\\
E-mail: zenchuk@itp.ac.ru
\\\\
P.M. Santini\\Dipartimento di Fisica, Universit\`a di Roma "La Sapienza" \\
and Instituto Nazionale di Fisica Nucleare, Sezione di Roma1, \\
Piazz.le Aldo Moro 2, I-00185 Roma, Italy\\
E-mail: paolo.santini@roma1.infn.it
}

\maketitle

\begin{abstract}
In this paper we construct nonlinear partial differential equations in more than 3 independent variables,  
possessing a manifold of analytic solutions with high, 
but not full, dimensionality. For this reason we call them ``partially integrable''. 
Such a construction is achieved using a suitable modification of the classical 
dressing scheme, consisting in assuming that the kernel of the basic integral operator of the dressing 
formalism be nontrivial. This new hypothesis leads to the construction of: 1) a linear system of compatible 
spectral problems for the solution $U(\lambda;x)$ of the integral equation 
in 3 independent variables each (while the usual dressing method generates spectral problems in 1 or 2 dimensions);  
2) a system of nonlinear partial differential equations in $n$ dimensions ($n>3$), possessing a manifold of analytic solutions 
of dimension ($n-2$), which includes one largely arbitrary relation among the fields. These nonlinear equations can also 
contain an arbitrary forcing.   

\end{abstract}

\section{Introduction}
\label{Intr}
Since the discovery of the integrability of 
the Korteweg-de Vries equation \cite{GGKM}, 
much effort has been devoted to the study of direct techniques
 to construct and solve nonlinear 
Partial Differential Equations (PDEs). One of the most 
powerful of such techniques is the dressing method, 
originally introduced in \cite{ZSh1} and subsequently generalized in \cite{ZSh2}-\cite{Zak} (see also \cite{ZMNP,K}), 
which is based 
on the existence of a linear analyticity problem, i.e. a Riemann-Hilbert or a $\bar\partial$ problem 
in some spectral variable $\lambda$ for some matrix function $U(\lambda;x)$,
 depending parametrically on the space-time 
variables $x=(x^1,\dots,x^n)$. (The $\bar\partial$ problem was introduced, in the context of integrable systems, 
in \cite{BC,ABF}.) Such an analyticity 
problem is characterized by a linear integral equation, 
whose unique solvability allows one to 
construct and solve an overdetermined system of 
compatible linear spectral problems for $U(\lambda;x)$, 
and, consequently, a nonlinear system of PDEs in 
the independent variables $x$, for the coefficients of such a linear system. 

The manifold of the analytic solutions of 
the nonlinear PDEs constructed by the dressing 
method is parameterized in terms of a proper number of arbitrary spectral 
functions, appearing in the 
linear integral equation, which depend on $n-1$ variables. Therefore  
the solution space is full and we say that the nonlinear PDE is 
completely integrable. For instance, the 
solution space of $1+1$ dimensional scalar systems like 
the Korteweg-de Vries and the nonlinear Schr\"odinger equations 
has dimension 1, being parameterized by an arbitrary function 
of 1 variable, while the 
solution space of the $2+1$ dimensional generalizations of them, 
the Kadomtsev-Petviashvili and the 
Davey-Stewartson equations, has dimensionality 2, being parameterized 
by an arbitrary function of 2 variables.

Motivated by the above considerations, in this paper we say that the dimensionality of the space of analytic solutions 
of a system of PDEs is $k$, if the analytic solutions are parameterized by a ``sufficient number'' of 
arbitrary functions of $k$ independent variables. For instance, if the system of PDEs contains  $K$ equations involving 
first order "time"-derivatives of $K$ functions, then the ``sufficient number'' is $K$. 
If $k=n-1$, then the systems is completely integrable. 

One of the most outstanding open problems in the theory of 
integrable systems is 
the construction of nonlinear PDEs in multi-dimensions (see, for instance, \cite{Zakharov1,Zakharov2}), 
i.e., in more than 3 dimensions, which could be integrated 
using suitable extensions of the above dressing procedure. 
Apart from few exceptional instances, 
among which one counts the  self-dual Yang-Mills equations 
\cite{YM} and the Plebanski 
heavenly equation \cite{Pleb} (see \cite{BPST,BZ2,ADHM,DM} 
and \cite{BK,MS} for their integration schemes), no 
significant examples are known 
in the literature \cite{AC}.

The purpose of this paper is the construction of PDEs 
in more than 3 independent variables,  
possessing a manifold of analytic solutions with high, 
but not full, dimensionality. For this reason we call 
such PDEs ``partially integrable''. 
This construction is achieved using a suitable 
modification of the classical 
dressing scheme, consisting in assuming that the 
kernel of the basic integral operator of the dressing 
formalism be nontrivial. As we shall see, this new 
hypothesis leads to the construction of: 
\begin{enumerate}
\item a linear system of compatible spectral problems in 3 independent variables each,     
for the eigenfunction $U(\lambda;x)$, where $\lambda$ is a {\it vector} spectral parameter 
(while the usual dressing method generates spectral problems in 1 or 2 dimensions with scalar spectral parameter); 
\item partially integrable nonlinear PDEs in $n$ dimensions, 
possessing a solution space of dimension ($n-2$). 
\end{enumerate}

A prototype example is given by the following 4 dimensional system of two matrix equations 
\begin{eqnarray}\label{Sec1:equation_proto}
 {\cal B}_2(q_1,q_1,q_2){\cal B}^{-1}_2(q_1,q_2,q_3)={\cal B}_3(q_1,q_1,q_2){\cal B}^{-1}_3(q_1,q_2,q_3)=
{\cal B}_4(q_1,q_1,q_2){\cal B}^{-1}_4(q_1,q_2,q_3)
 \end{eqnarray}
for the three square matrix fields $q_1(x),q_2(x),q_3(x)$, supplemented by the ``largely arbitrary'' relation
\begin{eqnarray}\label{Sec1:condition_proto}
F(q_1,q_2,q_3) = 0
\end{eqnarray}
among them, where the matrix blocks ${\cal B}_j$ are defined as: 
\begin{equation}\label{Sec1:def_proto}
{\cal B}_j(q_1,q_2,q_3)\equiv {q_2}_{x^j}-{q_2}_{x^1}B_j-q_2[q_1,B_j]-[B_j,q_3],\;\;\;j=2,3,4,
\end{equation}
$B_j,\;j=2,3,4$ are constant diagonal matrices different from the identity, and $[\cdot,\cdot ]$ is the usual 
commutator of matrices. In the simplest case, the largely arbitrary relation 
(\ref{Sec1:condition_proto}) can be chosen to be an equation defining one of the fields, say $q_3$, to be any given 
function $\gamma(x)$ (in general, a generalized function), interpretable as an ``external arbitrary forcing'': 
\beq
\label{Sec1:relation_proto}
F:\;\;\;\;\;q_3(x)=\gamma(x).
\eeq
As we shall see in the following, equations (\ref{Sec1:equation_proto}-\ref{Sec1:relation_proto}) possess a manifold of analytic 
solutions of dimension $2$.

The above closed system of equations (\ref{Sec1:equation_proto}-\ref{Sec1:condition_proto}) share with the 
other models constructed in this paper the following properties. 

\begin{enumerate}
\item The existence of a nontrivial kernel of the 
basic integral equation implies that the solutions constructed by the dressing depend on an arbitrary function $f(x)$ of the 
coordinates; this fact has the following  important implications. 
\item The nonlinear system of PDEs constructed by the dressing scheme possesses a distinguished 
block structure (see (\ref{Sec1:equation_proto})) and is under-determined. 
\item
To close the system and fix its indeterminacy (or, equivalently, to fix 
$f(x)$), one has to introduce an ``external and largely arbitrary'' relation among the fields (see (\ref{Sec1:condition_proto})).
If, for instance, such a relation (algebraic or differential) is linear, then the construction of explicit solutions via 
the dressing algorithm remains  
linear as well. The simplest example of linear relation is obtained imposing that one of the fields be a given function of the 
coordinates, like in (\ref{Sec1:relation_proto}), interpretable as an external forcing.  
\item
 The system of PDEs depends 
on two types of matrix fields, those obtained ``saturating the vector parameter $\lambda$'' of the solution $U(\lambda;x)$ 
of the linear integral equation by 
ingredients of the classical dressing method, whose dimensionality is constrained, and those 
obtained saturating $\lambda$ by a novel dressing function $G(\lambda;x)$, whose dimensionality is not constrained. That's why 
the dimensionality of the solution space, ($n-2$), can be arbitrarily large. 
\item
 While integrable PDEs in low dimensions (2+1 or less) are the compatibility of overdetermined systems of  
linear spectral problems, such a feature seems to be lost for our higher dimensional examples. 
\end{enumerate}

We remark that partially integrable equations of the type (\ref{Sec1:equation_proto}) are somehow connected to 
the N-wave type systems; indeed the two equations 
\begin{eqnarray}\label{Sec1:equation_Nwave}
 {\cal B}_2(q_1,q_1,q_2)={\cal B}_3(q_1,q_1,q_2)=0
\end{eqnarray}
are equivalent to the N-wave system in 2+1 dimensions for the field $q_1$, obtained eliminating $q_2$ from 
equations (\ref{Sec1:equation_Nwave}) (see also Sec. \ref{Section:class}). The construction 
of partially integrable PDEs connected to other basic integrable systems, like those 
belonging to the Kadomtsev-Petviashvili hierarchy, or those associated with the   
Davey-Stewartson equation, will be the subject of a forthcoming paper.  

We also remark that a different generalization of the dressing procedure, allowing to construct a class of partially integrable 
PDEs which combine $S$ and $C$ integrability features, has been already proposed in \cite{Z} 
(a nonlinear system is $S$ integrable, like the KdV equation, if it is solved via a linear 
integral equation; it is $C$ integrable, like the Burgers equation, if it is linearized by a simpler change of variables, 
like a contact transformation; 
see \cite{Calo} for more details on these definitions).

The paper is organized as follows. In Sec. \ref{Section:Classical} we derive, for the sake of comparison, 
the well-known $N$-wave system in (2+1)-dimensions, using the classical dressing method, 
 and we show that its solution space is full.    
Then in Sec. \ref{Section:Nw} we explore the implications of the existence of a nontrivial kernel of the 
basic integral operator of the dressing scheme, and we construct examples 
of nonlinear $n$-dimensional PDEs possessing a space of analytic solutions of dimension $n-2$. We also show 
that these equations do not seem to be the commutativity condition of overdetermined systems 
of linear PDEs. In Sec. \ref{Section:Solution} we show how to construct an integral operator possessing a nontrivial 
kernel, and we use this result to characterize a large class of analytic solutions of the partially integrable PDEs 
of this paper. A natural extension of the algorithm presented in this paper is briefly mentioned in the final  
 Sec. \ref{Section:conclusion}.

\section{Derivation of classical integrable systems using the  
dressing algorithm}
\label{Section:Classical}

The starting point of the dressing constructions contained in this paper is the linear integral 
equation
\begin{eqnarray} \label{Sec1:U}
\Phi(\lambda;x)=\int \Psi(\lambda,\mu;x)U(\mu;x)d\Omega(\mu)
\equiv \hat\Psi U(\lambda;x),
\end{eqnarray}
in the spectral variables $\lambda$, $\mu$, 
for the unknown matrix function $U$. The given matrix functions $\Phi$ and $\Psi$ are defined by 
some extra conditions, which fix their dependence on an additional vector parameter $x= (x^1,\dots,x^{n})$, 
whose components are the independent variables of the associated nonlinear PDEs. $\Omega$ is some largely arbitrary 
scalar measure in the $\mu$-space. Apart from $\Omega$, all the functions appearing in this paper 
are $Q\times Q$ matrix functions.

We remark that no a priori assumption is made in (\ref{Sec1:U}) 
on the dependence of $\Psi$ on $\lambda$ 
(this general starting point has been used, for instance, in 
\cite{SAF} and in \cite{Z}), to keep the structure 
of $\Psi$ as much general as possible. Indeed, although in 
most of the cases such a dependence is described by a Cauchy kernel, an  
indication that equation (\ref{Sec1:U}) is the manifestation of 
Riemann-Hilbert and/or $\bar\partial$ analyticity problems, 
there are examples (see \cite{SAF} and \cite{Z}) in which       
more general representations appear, 
indicating that the above analyticity 
problems could be a too restrictive starting points.

Before developing, in Sec. \ref{Section:Nw}, \ref{Section:Solution}, the novel features of the dressing method, it is useful to 
summarize the essential steps of the classical dressing method used to construct and solve 
the classical 3-dimensional $N$-wave system (\ref{Sec0:Nw1}) (which is known to be  
an $S$-integrable system), together with its solution space. Such a solution space is 2 dimensional 
(i.e., it is complete), being parameterized by 
an arbitrary function of 2 variables. 

\subsection{$S$-integrable PDEs: the $N$-wave system.}
\label{Section:class}

The basic assumption underlying all the known dressing procedures available in the literature, is that the operator $\hat\Psi$ 
in (\ref{Sec1:U}) be uniquely invertible; i.e., that
\begin{eqnarray}\label{Sec1:nondeg}
\dim {\mbox{ker}} \hat\Psi =0.
\end{eqnarray}
The $x$-dependence is introduced by the matrix equations 
\begin{eqnarray} \label{Sec1:x}
\Psi_{x^i}(\lambda,\mu;x)=\Phi(\lambda;x)B_iC(\mu;x),~~~i=1,..,{\mbox{dim}}~x,
\end{eqnarray}
showing that the $x$-derivatives of the kernel $\Psi$ are degenerate matrix 
functions of the spectral parameters, another basic feature 
of all known dressing algorithms,   
where $B_i,~i=1,..,n,$ are  constant diagonal matrices, so at
most $Q$ of them may be independent. Due to the above degeneracy, the compatibility 
of equations (\ref{Sec1:x}) leads to separate equations for $\Phi$ and $C$: 
\begin{eqnarray} \label{Sec1:Phi_x}
\Phi_{x^i} B_j - \Phi_{x^j} B_i = 0,~~~~i\ne j,\\\label{Sec1:c_x}
B_j C_{x^i} - B_i C_{x^j} = 0,~~~~i\ne j,
\end{eqnarray}
and one equation is the adjoint of the other. Without loss of generality we assume $B_1=I$, where $I$ is the identity matrix.

Replacing, in equation (\ref{Sec1:Phi_x}), $\Phi$ by $\hat\Psi U$, as indicated in (\ref{Sec1:U}), and using 
(\ref{Sec1:x}), one obtains the following equation:
\begin{eqnarray}
\label{Sec1:PsiU}
\hat\Psi L_{ij}U=0, 
\end{eqnarray}
where
\begin{eqnarray}
\label{Sec1:LUU}
L_{ij}U\equiv U_{x^i} B_j - U_{x^j} B_i+ UB_i v B_j - U B_j v B_i,~~i,j=1,..,{\mbox{dim}}~x,~~~i\ne j
\end{eqnarray}
and 
\begin{eqnarray}
\label{Sec1:v}
v(x)\equiv \int C(\lambda;x)U(\lambda;x)d\Omega(\lambda).
\end{eqnarray}
Then the property (\ref{Sec1:nondeg}) implies that 
 \begin{eqnarray}\label{Sec0:UU}
 L_{ij} U(\lambda;x)=0,~~i,j=1,..,{\mbox{dim}}~x,~~~i\ne j
 \end{eqnarray}
 or, explicitly:
\begin{eqnarray}\label{Sec0:U_lin}
L_{21}U=U_{x^2}   - U_{x^1} B_2 - U [v, B_2]=0,\\\nonumber
L_{31}U=U_{x^3}   - U_{x^1} B_3 - U [v, B_3]=0,\\\nonumber
\end{eqnarray}
having chosen $j=1$, $i=2, 3$. 

This is nothing but the well-known linear overdetermined 
system corresponding to the $N$-wave equation in the three variables $x^1,x^2,x^3$. 

The associated complete system of nonlinear PDEs is simply obtained, in the dressing philosophy, upon
``saturating the parameter $\lambda$'' in equations (\ref{Sec0:U_lin}) by the integral operator 
$\int d\Omega(\lambda) C(\lambda;x)\cdot$:
\begin{eqnarray}\label{Sec0:nl}
L_{21} v-[B_2,v^1]=v_{x^2} - v_{x^1} B_2 - v [v, B_2] 
-[B_2, v^1]  =0,\\\nonumber
L_{31} v-[B_3,v^1]=v_{x^3} - v_{x^1} B_3 - v [v, B_3] 
-[B_3, v^1]  =0.
\end{eqnarray}
It is written in terms of the square matrix fields $v(x)$ and $v^1(x)$, where
\begin{eqnarray}\label{Sec0:v1}
v^1(x)\equiv \int C_{x^1}(\lambda;x)U(\lambda;x)d\Omega(\lambda).
\end{eqnarray}
Eliminating $v^1$ from these two equations, we get the celebrated $N$-wave system in $3$ dimensions:
\begin{eqnarray}\label{Sec0:Nw1}
[v_{x^3},B_2] - [v_{x^2},B_3] + B_2 v_{x^1} B_3 -  B_3 v_{x^1} B_2
-[[v,B_2],[v,B_3]]=0.
\end{eqnarray}
The same equation may be derived directly from the compatibility condition of the system
(\ref{Sec0:U_lin}).

Similarly, considering 
the equations $L_{j1}U=0$ and $L_{k1}U=0$ for any $ j \neq k \neq 1$, one  
derives the hierarchy of $n$-wave equations
 \begin{eqnarray}\label{Sec0:Nw2}
[v_{x^k},B_j] - [v_{x^j},B_k] + B_j v_{x^1} B_k -  B_k v_{x^1} B_j
-[[v,B_j],[v,B_k]]=0.
 \end{eqnarray}

We remark that, in the above dressing construction, the linear integral operator $\hat\Psi$ in (\ref{Sec1:U}) 
acts from the left and, consequently, the partial differential operators $L_{ij}$ in (\ref{Sec0:UU}) 
act from the right, while, in the soliton literature, one usually makes the opposite choice. 

Our choice is motivated by the fact that, as we shall see in Sec.\ref{Section:Nw},\ref{Section:Solution},  
in more than $2+1$ dimensions, the role played by the linear integral 
equation (\ref{Sec1:U}) seems to be more 
fundamental  than that played by the linear overdetermined system of PDEs. Indeed,  
while integrable PDEs in 2+1 dimensions (or less) are characterized as the compatibility condition of 
a linear overdetermined system of PDEs, such a basic property seems to be lost in multi-dimensions. Instead, as we 
shall see in the following sections, the linear integral equation (\ref{Sec1:U}) can generate nonlinear PDEs, 
together with their large  
manifold of analytic solutions, also in the multidimensional context.
 
\subsection{Solution space}
 
We now consider the manifold of particular solutions of 
equations (\ref{Sec0:Nw1},\ref{Sec0:Nw2}).
The solutions of eqs.(\ref{Sec1:Phi_x}) and (\ref{Sec1:c_x}) can be parameterized as follows:
\begin{eqnarray} \label{Sec0:Phi}
\Phi(\lambda;x)= \int \Phi_0(\lambda,k)e^{kB\cdot x} d k,\\
\label{Sec0:c}
C(\mu;x)=\int e^{qB\cdot x} C_0(\mu,q) d q,
\end{eqnarray}
where
\begin{equation}
B\cdot x=\sum_{i=1}^{n}B_i x^i,
\end{equation}
and where the spectral parameters $\lambda,\mu, k,q$ are necessarily scalars. 
Thus equations (\ref{Sec1:x}) yield:
\begin{eqnarray}\label{Sec2:Psi}
\Psi(\lambda,\mu;x)= 
\int \Phi_0(\lambda,k)e^{(k+q)B\cdot x} 
C_0(\mu,q)\frac{dk dq }{k+q}+\Sigma(\lambda,\mu),\;\;\;\;\;B_1=I, 
\end{eqnarray}
where the integration constant $\Sigma(\lambda,\mu)$ is 
chosen here to be $\delta(\lambda-\mu)$.
 
It is simple to see, from the linear limit,  that the solution space of equation (\ref{Sec0:Nw1}), generated 
by the dressing algorithm, is full. Indeed, in the linear limit: 
$\Psi(\lambda,\mu) \sim \delta(\lambda-\mu)$ and $U\sim \Phi$. Take 
$C_0(\lambda,q) = \delta (\lambda- q)$;  
then the solution $v$ of the 3 - dimensional $N$-wave system (\ref{Sec0:Nw1}), which in the linear limit reads 
\begin{eqnarray}
\label{dim2}
v(x)\sim \int C(\lambda;x)\Phi(\lambda;x) d\Omega(\lambda) =
 \int e^{\lambda B\cdot x} \Phi_0(\lambda,k)e^{kB\cdot x} d kd\Omega(\lambda),
\end{eqnarray}
is parameterized by the arbitrary matrix function $\Phi_0(\lambda,k)$ of the two scalar spectral 
parameters $\lambda,k$; then its solution space is 2 dimensional, and therefore it is complete. 

We end this section remarking that the Cauchy kernel appearing 
in (\ref{Sec2:Psi}), obtained here as a consequence of 
equations (\ref{Sec0:Phi}),(\ref{Sec0:c}) and (\ref{Sec1:x}), 
is a manifestation of the distinguished analyticity 
properties of the solution $U(\lambda;x)$ in the complex $\lambda$ plane, 
in agreement with the well-known derivations of the $N$-wave equation  (\ref{Sec0:Nw1}) 
from Riemann-Hilbert and /or $\bar\partial$ problems \cite{Kaup,F,FA}.

\section{Partially integrable PDEs in multi-dimensions.}
 \label{Section:Nw}
In this section we show how to construct partially integrable PDEs in $n$ dimensions exhibiting 
a space of analytic solutions of dimension $(n-2)$.

\subsection{Generalization of the dressing algorithm}
\label{dressing}
In the Sec.\ref{Section:class}  we have constructed, 
from the general hypothesis (\ref{Sec1:U}), (\ref{Sec1:nondeg}) 
and (\ref{Sec1:x}) 
underlying the classical dressing algorithm, 
the integrable $N$-wave system in $3$ dimensions. 
The main obstacle to go to higher dimensions  
is clearly due to the fact that each linear problem  
$L_{ij}U=0$, as a consequence of (\ref{Sec1:nondeg}), is 2 dimensional. 

To increase the dimensionality of the 
linear problems, we then suppose that the kernel of the operator $\hat\Psi$ is one dimensional:
\begin{eqnarray}\label{Sec2:deg}
\dim\mbox{ker}\hat\Psi =1;
\end{eqnarray}
i.e., the solution of the homogeneous equation associated with eq.(\ref{Sec1:U}) is nontrivial:
\begin{eqnarray}\label{Sec1:Unhom}
 0=\hat\Psi H \;\;\Leftrightarrow \;\; H(\lambda;x) = U^h(\lambda;x)f(x),
\end{eqnarray}
where $U^h(\lambda;x)$ is some nontrivial solution of the homogeneous equation $\hat\Psi H=0$,
$f(x)$ is an arbitrary matrix function of $x$, and $\lambda$, $\mu$, $\nu$ are now   
vector spectral parameters whose dimension is specified in Sec.\ref{Section:Solution}. 
Then the general solution of eq.(\ref{Sec1:U}) reads
\begin{eqnarray}\label{Sec1:Uninhom}
U(\lambda;x)=U^p(\lambda;x) + U^h(\lambda;x) f(x),
\end{eqnarray}
where $U^p(\lambda;x)$ is some particular solution of (\ref{Sec1:U}).

As a consequence of the novel assumption (\ref{Sec2:deg}), equation (\ref{Sec1:PsiU}) implies the following equations
 for $U$:
 \begin{eqnarray}\label{Sec1:LU0}
 L_{ij} U(\lambda;x) = (L_{nm} U(\lambda;x)) A^{ijnm}(x),~~~i\ne j,~~n\ne m,
 \end{eqnarray}
where $A^{ijnm}(x)$ are some matrix functions of $x$ to be specified, reflecting the fact that two solutions of the homogeneous 
equation $\hat\Psi H=0$ differ by a matrix function of $x$, multiplied from right.

Since the following cyclic permutation formula among three operators $L_{ij}$ holds:
\begin{eqnarray}
\sum_{cycl(ijk)} (L_{ij} {\cal{M}}) B_k = 0,~~~i\ne j\ne k,
\end{eqnarray}
where ${\cal{M}}$ is an arbitrary square matrix, it follows that only ($n -1$) operators $L_{ij}$ are linearly independent.  
Therefore we take the operators $\{L_{j1}$, $j=2,\dots, n\}$ as elements of the basis, 
and we consider the following subset of equations (\ref{Sec1:LU0}), involving only these elements:
\begin{eqnarray}\label{Sec1:LU}
{\cal{E}}_j(\lambda;x)\equiv  L_{j1} U(\lambda;x) - (L_{21} U(\lambda;x)) A^{j}(x)=0,\;\;\;j=3,\dots, n, \\
L_{j1}U\equiv U_{x^j} -U_{x^1} B_j -U [v,B_j],\;\;\;\;j=2,\dots,\mbox{dim }x,
\end{eqnarray}
where $A^j(x)$ are some matrix functions to be defined. 

We have established that, if $\dim\mbox{ker}\hat\Psi =1$, then each 
linear equation (\ref{Sec1:LU}) for the spectral function $U(\lambda;x)$ is 3 dimensional.  

The associated nonlinear equations, obtained ``saturating the parameter 
$\lambda$'' in equations (\ref{Sec1:LU}) by 
the integral operator $\int d\Omega(\lambda)  C(\lambda;x)\cdot$, read
\begin{equation}\label{Sec1:nl_j0}
L_{j1}v-[B_j,v_1]-(L_{21}v-[B_2,v^1])A^j=0,\;\;\;j=3,.., n;
\end{equation}
they are given in terms of the fields $v(x)$ and $v^1(x)$, defined respectively in (\ref{Sec1:v}) and (\ref{Sec0:v1}), and of 
the matrices $A^j(x)$.  More explicitly, one obtains:
\begin{equation}\label{Sec1:nl_j}
\begin{array}{l}
v_{x^j} - v_{x^1} B_j - v [v, B_j] -[B_j, v^1]-\left(v_{x^2} - v_{x^1} B_2 - v [v, B_2] 
-[B_2, v^1]\right) A^j=0,\;\;j=3,.., n.
\end{array}\end{equation}

In order to express $A^{j}(x)$ in terms of $U$ and close the system, we introduce 
an external dressing function $G(\lambda;x)$, and the associated new matrix fields 
\begin{eqnarray}\label{Sec0:w^ij}
w^{00}(x)\equiv \int G(\lambda;x)U(\lambda;x)d\Omega(\lambda),\;\;
w^{j0}(x)\equiv \int G_{x^j}(\lambda;x)U(\lambda;x)d\Omega(\lambda),\;\;j>0, \\
w^{ij}(x)\equiv \int G_{x^ix^j}(\lambda;x)U(\lambda;x)d\Omega(\lambda),\;\;i,j>0,\;\;
w^{ij}(x)=w^{ji}(x).
\end{eqnarray}

The equations for the fields $w^{ij}$ can be derived applying
 $\int d\Omega(\lambda) G(\lambda) \cdot$ and
  $\int d\Omega(\lambda) G_{x^n}(\lambda) \cdot$ 
  to the linear equation (\ref{Sec1:LU}), obtaining: 
 \begin{eqnarray}\label{Sec1:gU}
 L_{j1}w^{n0}  - w^{jn} + w^{1n} B_j =
 (L_{21}w^{n0}  - w^{2n} +  w^{1n} B_2) A^j,~j=3,..,\mbox{dim}x,~n=0,1,..,\mbox{dim}x.
 \end{eqnarray}
 
Some of these equations can be taken as definition of $A^j$. 
 But, to close the system, one needs to introduce 
 the following additional structures.

\noindent
(a) Equations defining $G(\lambda;x)$. Since $G$ is an outer dressing function, these 
equations can be a quite arbitrary system of compatible and solvable 
 partial differential equations, either linear or nonlinear, 
involving derivatives of any order and dimension. This freedom plays a key role in 
allowing for a large solution space. 

\noindent
(b) An additional relation between all the matrix fields, which may be taken in quite arbitrary form
\begin{eqnarray}\label{Sec1:condition00}
F(v,v^1,w^{00},w^{i0},w^{ij})=0,\;\;\;\;i,j=1,2,\dots.
\end{eqnarray}
This equation is needed to fix the arbitrary function $f(x)$ of the 
 variables $x^i$ appearing in the solution of the inhomogeneous equation (\ref{Sec1:U}) 
(see equations (\ref{Sec1:Unhom},\ref{Sec1:Uninhom})).
The relation (\ref{Sec1:condition00}) is largely arbitrary; the only requirement 
 is that it should give rise to a solvable equation  
 for $f(x)$.  
 Since $U$ depends linearly on the arbitrary function $f(x)$,
 all the fields depend linearly on $f$ as well. Indeed, using
 equations (\ref{Sec1:v}, \ref{Sec0:v1}, \ref{Sec0:w^ij}) and equation (\ref{Sec1:Uninhom}),
 one obtains
 \begin{eqnarray}\label{Sec1:fields_f}
 v(x)&=&h^v_0(x)+h^v_1(x) f(x),\;\;h^v_0(x=\int C(\lambda) U^p(\lambda) d\Omega(\lambda),\;\;
 h^v_1=\int C(\lambda) U^h(\lambda) d\Omega(\lambda),\;\;\\\nonumber
 v^1(x)&=&h^{v^1}_0(x)+h^{v^1}_1(x) f(x),\;\;h^{v^1}_0=\int C_{x^1}(\lambda) 
 U^p(\lambda) d\Omega(\lambda),\;\;
 h^{v^1}_1=\int C_{x^1}(\lambda) U^h(\lambda) d\Omega(\lambda),\;\;\\\nonumber
w^{00}(x)&=&h^{w^{00}}_0(x)+h^{w^{00}}_1(x) f(x),\;\;h^{w^{00}}_0=
 \int G(\lambda) U^p(\lambda) d\Omega(\lambda),\;\;
 h^{w^{00}}_1=\int G(\lambda) U^h(\lambda) d\Omega(\lambda),\;\;\\\nonumber
 w^{i0}(x)&=&h^{w^{i0}}_0(x)+h^{w^{i0}}_1(x) f(x),\;\;h^{w^{i0}}_0=
 \int G_{x^i}(\lambda) U^p(\lambda) d\Omega(\lambda),\;\;
 h^{w^{i0}}_1=\int G_{x^i}(\lambda) 
 U^h(\lambda) d\Omega(\lambda),\\\nonumber
 w^{ij}(x)&=&h^{w^{ij}}_0(x)+h^{w^{ij}}_1(x) f(x),\;\;h^{w^{ij}}_0=
 \int G_{x^ix^j}(\lambda) U^p(\lambda) d\Omega(\lambda),\;\;
 h^{w^{ij}}_1=\int G_{x^ix^j}(\lambda) 
 U^h(\lambda) d\Omega(\lambda),
 \end{eqnarray}
 where all the $h$'s are known functions of $x$, and $i,j=1,2,\dots$.
  Using this fact, the following types of relation (\ref{Sec1:condition00}) open different scenarios:
 \begin{enumerate}
 \item
 $F$ is an algebraic expression of its arguments, leading
  to an algebraic equation for $f$.
 The simplest case is, of course, that of a linear equation, leading to a linear algebraic equation 
for $f(x)$. The simplest example of linear algebraic relation is obtained imposing that one of the fields 
be a ``given function of the coordinates, interpretable as an external forcing''.   
 \item
 $F$ is a multidimensional linear partial differential equation of any dimension and order, either with constant or variable
 coefficients. This leads to a linear PDE for $f(x)$ having the same dimension and
 order, and always  variable coefficients (due to the functions $h$ in equations (\ref{Sec1:fields_f})).  
\item
$F$ is a nonlinear PDE, whose dimension $m$ is lower than the dimensionality $n$ of the  
system of PDEs. This leads to a nonlinear PDE for $f(x)$ in $m$ dimensions, but with variable coefficients. 
In this case, the dressing procedure allows one to replace the nonlinear PDEs under investigation 
by a nonlinear PDE for $f(x)$ in lower dimensions (a ``reduction of complications'').  
  
 \end{enumerate}
 
Among all these cases, the most remarkable one is when $F$ is a linear algebraic
 relation, since $f(x)$ can be found explicitly and the solution manifold may be constructed analytically 
(see Sec.\ref{Section:Fixing} for more details on this point). 

The equation defining $G(\lambda;x)$, together with 
the relation (\ref{Sec1:condition00}) among the fields, 
provides the completeness of the nonlinear system of PDEs for 
the fields $v,~v^1$ and $w^{ij}$ generated by the dressing. 

\subsection{Examples}
\label{Section:examples}

In this section we consider some basic examples of partially integrable PDEs, corresponding to special 
 definitions of $G(\lambda;x)$, and to particular relations (\ref{Sec1:condition00}).

\subsubsection{The simplest nonlinear partially integrable PDEs}
\label{Section:simplest}

The simplest possible case corresponds to a function $G$ independent of $x$. Then we 
have the only additional field $w^{00}$, since $w^{ij}=0$, $i,j>0$. We consider two examples 
of relation (\ref{Sec1:condition00}). 

\noindent
{\bf 1.} The relation (\ref{Sec1:condition00}) is chosen as follows:
\begin{eqnarray}\label{Sec1:condition0}
F:~~~~~w^{00}(x)=\exp\left[\sum_{i=1}^{n} {a_i x^i}\right],
\end{eqnarray}
where $a_j$ are constant diagonal matrices.
Then equation (\ref{Sec1:gU}) with $n=0$ yields:
\begin{eqnarray}
A^j=\big(a_2-a_1 B_2 +[B_2,v]\big)^{-1}\big(a_j-a_1 B_j+[B_j,v]\big).
\end{eqnarray}
Consequently, equations (\ref{Sec1:nl_j}), for each particular choice of $j$, involve just the two
fields $v$ and $v^1$. Thus we need two equations of this type to close the
system, say $j=3,4$, obtaining the following system of 2 matrix equations in 4 dimensions for 
the matrix fields $v,v^1$:
\begin{equation}
\label{ex1}
\begin{array}{l}
\big(L_{21}v-[B_2,v_1]\big)\big(a_2-a_1 B_2 +[B_2,v]\big)^{-1}=\\
\big(L_{31}v-[B_3,v_1]\big)\big(a_3-a_1 B_3 +[B_3,v]\big)^{-1}=
\big(L_{41}v-[B_4,v_1]\big)\big(a_4-a_1 B_4 +[B_4,v]\big)^{-1}.
\end{array}
\end{equation}
{\bf 2.} The relation (\ref{Sec1:condition00}) is chosen as follows:
\begin{eqnarray}\label{Sec1:condition01}
F:~~~~~v^1(x)=\gamma(x),
\end{eqnarray} 
where $\gamma(x)$ is an arbitrary matrix function of $x$. In this case we choose:
\beq
A^j(x)=\left(L_{21}w^{00}(x)\right)^{-1}L_{j1}w^{00}(x),~~j=3,4,
\eeq 
and we obtain the following system of 2 matrix equations in 4 dimensions for 
the two matrix fields $v,w^{00}$:
\beq
\label{ex2}
\ba{l}
\left(L_{21}v(x)-\gamma_2(x)\right)\left(L_{21}w^{00}(x)\right)^{-1}=
\left(L_{31}v(x)-\gamma_3(x)\right)\left(L_{31}w^{00}(x)\right)^{-1}= \\
\left(L_{41}v(x)-\gamma_4(x)\right)\left(L_{41}w^{00}(x)\right)^{-1}, \\
~~ \\
\gamma_j(x)\equiv [B_j,\gamma(x)],~~~j=2,3,4,
\ea
\eeq
depending on the arbitrary forcing $\gamma(x)$. 

In section Sec. \ref{Section:Solution} we will see that the space of analytic solutions of these two systems is 2-dimensional.

\subsubsection{Partially integrable n-dimensional PDEs}
\label{Section:general} 
In this section we first derive nonlinear PDEs in 4 dimensions, and then we turn to nonlinear PDEs in arbitrary dimensions.

Let $G$ be defined by the following equations
\begin{eqnarray}
\label{g-x}
 G_{x^j}=\alpha_{j} G_{x^1},\;\;\;j=1,\dots,n,
 \end{eqnarray}
 where $\alpha_j$ are constant diagonal matrices. 

 The special form (\ref{g-x}) of $G$ implies that the fields $w^{j0}(x),\;w^{jn}(x)$ can be expressed in terms 
of the fields $w^{10}(x),\;w^{11}(x)$ respectively:
\begin{eqnarray}
w^{j0}(x)=\alpha_j w^{10}(x),\;\;\;w^{jk}(x)=\alpha_j\alpha_k w^{11}(x)\;\;\;\;j,k=1,\dots,n.
\end{eqnarray}
 Then all the equations (\ref{Sec1:nl_j}) and (\ref{Sec1:gU}) reduce to the 6 equations (for $j=3,4$):
 \begin{eqnarray}\label{Sec1:Eq_gen1}
 && L_{j1}v -[B_j,v^1] = (L_{21}v -[B_2,v^1]) A^j,\\\label{Sec1:Eq_gen2}
 && L_{j1}w^{00} -\alpha_j w^{10} + w^{10} B_j = 
  (L_{21}w^{00} -\alpha_2 w^{10} + w^{10} B_2) A^j,\\\label{Sec1:Eq_gen3}
 && L_{j1}w^{10} -\alpha_j w^{11} + w^{11} B_j = 
  (L_{21}w^{10} -\alpha_2 w^{11} + w^{11} B_2) A^j,
 \end{eqnarray}
 Equations (\ref{Sec1:Eq_gen3}) can be viewed, for instance, as defining the matrix fields $A^j,\; j=3,4$; 
substituting these definitions of $A^j$ in equations (\ref{Sec1:Eq_gen1}) and (\ref{Sec1:Eq_gen2}), we 
obtain 4 matrix equations for the 5 matrix fields $v,v^1,w^{00},w^{10},w^{11}$. Before performing such operations, 
it is convenient to introduce a more compact notation defining the matrices
 \begin{eqnarray}
 E^v_j:=L_{j1}(v) -[B_j,v^1], \;\;
  E^{w^{k0}}_j:=L_{j1}(w^{k0}) -\alpha_j w^{1k} + w^{1k} B_j,\;\;j=3,4,\;\;k=0,1. 
 \end{eqnarray}
 Then equation (\ref{Sec1:Eq_gen3}) yields
 \begin{eqnarray}\label{defA_D1}
 A^j=  (E^{w^{10}}_2)^{-1} E^{w^{10}}_j,\;\;j=3,4
 \end{eqnarray}
 and the nonlinear system (\ref{Sec1:Eq_gen1},\ref{Sec1:Eq_gen2}) takes the form 
 \begin{eqnarray}\label{Sec1:Eq_short1}
 E^{v}_j(E^{w^{10}}_j)^{-1}&=&E^{v}_2(E^{w^{10}}_2)^{-1},\;\;\;\;\;j=3,4,
 \\\label{Sec1:Eq_short2}
 E^{w^{00}}_j(E^{w^{10}}_j)^{-1}&=&E^{w^{00}}_2(E^{w^{10}}_2)^{-1},\;\;\;\;\;\;j=3,4.
 \end{eqnarray}
  
This 4 dimensional system is not closed, consisting of 4 equations for the 5 functions
  $v,v^1,w^{00},w^{10},w^{11}$. This indeterminacy is consistent with the fact that the above equations are 
generated by the linear integral equation (\ref{Sec1:U}), which possesses a 1 dimensional space of homogeneous solutions. 
Therefore all the solutions of equations (\ref{Sec1:Eq_short1}-\ref{Sec1:Eq_short2}), constructed by the dressing 
procedure, contain an arbitrary function $f(x)$.  

To fix this arbitrary function, we use the relation (\ref{Sec1:condition00}) which, for this example, reads:
 \begin{eqnarray}\label{Sec1:condition_g}
 F(v,v^1,w^{00},w^{10},w^{11}) = 0.
 \end{eqnarray}
 As we discussed above, 
 this relation is largely arbitrary; 
 the only requirement is that it should give rise to a solvable equation  
 for the function $f(x)$ (see Sec.\ref{Section:Fixing} for more details on this point). 

We remark that the arbitrary relation (\ref{Sec1:condition00}) can be chosen in order to 
put equations (\ref{Sec1:Eq_short1},\ref{Sec1:Eq_short2}) into a differential polynomial form. 
If we choose, for instance, the bilinear relation 
\begin{eqnarray}
\label{bilinear}
F:~~~~~E^{w^{10}}_2=L_{21}(w^{10}) - \alpha_2 w^{11} + w^{11} B_2 = T,
\end{eqnarray}
where $T$ is a constant matrix,  
multiplying equations (\ref{Sec1:Eq_short1},\ref{Sec1:Eq_short2}) from right by 
$E^{w^{10}}_j$, one transforms them into a differential polynomial form.  
In addition, if $B^2=0$, the bilinear relation becomes linear: $w^{10}_{x^2} - \alpha_2 w^{11}  = T$.

We have established that, due to the novel 
hypothesis $\dim {\mbox{ker}}\hat\Psi =1$, the dressing algorithm allows 
one to construct a system of partially integrable PDEs in multi-dimensions which includes one 
largely arbitrary  relation among the fields. This is a novel and surprising feature of the theory.
  
It is interesting to remark that equations (\ref{Sec1:Eq_short1}-\ref{Sec1:Eq_short2}) admit the reduction 
\beq
v=w^{00},~v^1=w^{10},
\eeq
which follows imposing that $G(\lambda;x)=C(\lambda;x)$, as a consequence of the choice $\alpha_j=B_j$. 
In this case, equations  
(\ref{Sec1:Eq_short1}-\ref{Sec1:Eq_short2}) reduce to the following two nonlinear PDEs     
\begin{eqnarray}\label{Sec1:Eq_short3}
 E^{w^{00}}_2(E^{w^{10}}_2)^{-1}=E^{w^{00}}_3(E^{w^{10}}_3)^{-1}=E^{w^{00}}_4(E^{w^{10}}_4)^{-1}
 \end{eqnarray}
for the three matrix fields $w^{00},w^{10},w^{11}$, supplemented by the (largely) arbitrary relation
\begin{eqnarray}\label{Sec1:condition2}
 F(w^{00},w^{10},w^{11}) = 0.
 \end{eqnarray}
A simple example of linear relation (\ref{Sec1:condition2}) is, for instance, 
\beq
\label{rel_gamma} 
F:\;\;\;\;\;w^{11}(x)=\gamma(x),
\eeq 
where $\gamma$ is an arbitrary function; then the 
nonlinear system is given by the two 
equations (\ref{Sec1:Eq_short3}) for the two fields $w^{00},w^{10}$, with the arbitrary forcing $\gamma(x)$ 
(it is the system (\ref{Sec1:equation_proto}-\ref{Sec1:relation_proto}) presented in the introduction with a different 
notation).

As we shall see in section \ref{Section:dimensionality}, the above nonlinear PDEs in 4 dimensions possess an analytic solution 
space of dimension 2.
 
One can generalize the above construction, to generate partially integrable PDEs in $n$ dimensions ($n\ge 4$), whose manifold of 
analytic solutions has dimension $(n-2)$. This higher dimensional generalization is 
associated with a more general equation for $G$. 

For instance, in order to derive partially integrable PDEs in $5$ dimensions, one chooses $G(\lambda;x)$ to be defined by the 
following equations: 
\begin{eqnarray}\label{Sec1:g_D2}
G_{x^j}=\sum_{k=1}^2 \alpha_{jk} G_{x^k} ,\;\; j>2,
\end{eqnarray}
where $\alpha_{jk}$ are constant diagonal matrices. This special form implies that the fields $w^{j0},w^{j1},w^{j2},~j>2$ 
can be expressed in terms of the fields  $w^{i0},w^{ik},~i,k=1,2$:
\begin{eqnarray}
w^{j0}=\sum_{s=1}^2 \alpha_{js}w^{s0},\;\;\;w^{jk}=\sum_{s=1}^2 \alpha_{js}w^{sk},\;\;j>2,\;\;k=1,2. 
\end{eqnarray}
Then the system (\ref{Sec1:nl_j},\ref{Sec1:gU}) reduces to the following 12 equations:
\begin{eqnarray}\label{Sec1:Eq_gen1_v}
&& L_{j1}v -[B_j,v^1] = (L_{21}v -[B_2,v^1]) A^j,
  \;\;j=3,4,5,\\\label{Sec1:Eq_gen1_wj0}
 && L_{j1}w^{k0} - \sum_{i=1}^2 \alpha_{ji} w^{ki} + w^{k1} B_j = 
  (L_{21}w^{k0} - w^{k2} + w^{k1} B_2) A^j,\;j=3,4,5,\;k=0,1,2.
\end{eqnarray}
Following the previous procedure, we use equations 
(\ref{Sec1:Eq_gen1_wj0}) for $k=1$ to define the matrices $A^j$, and we 
substitute these definitions into 
equations (\ref{Sec1:Eq_gen1_v}) and (\ref{Sec1:Eq_gen1_wj0}) for $k=0$. Defining the blocks:
\begin{eqnarray}
E^{v}_j &\equiv& L_{j1}v -[B_j,v^1],\;\;j=2,3,4,5,\\\nonumber
E^{w^{k0}}_j &\equiv& L_{j1}w^{k0} - \sum_{i=1}^2 
\alpha_{ji} w^{ik} + w^{1k} B_j,\;
\;j=3,4,5,\;\;k=0,1,2,\\\nonumber
E^{w^{k0}}_2 &\equiv&  L_{21}w^{k0} - w^{k2} + w^{k1} B_2,\;\;k=0,1,2,
\end{eqnarray}
we obtain:
\begin{eqnarray}\label{defA_D2}
A^j=(E^{w_{10}}_2)^{-1} E^{w_{10}}_j,\;\;j=3,4,5.
\end{eqnarray}
Substituting (\ref{defA_D2}) into (\ref{Sec1:Eq_gen1_v}) and (\ref{Sec1:Eq_gen1_wj0}), we obtain the following 
closed system of 8 matrix PDEs for 
the 8 matrix fields $v,v_1,w_{00},w_{10}$, $w_{20},w_{11},w_{12},w_{22}$, 
consisting of the following 5 basic equations
\begin{eqnarray}\label{D=2basic}
  E^{v}_j(E^{w^{10}}_j)^{-1}&=&E^{v}_2(E^{w^{10}}_2)^{-1},\;\;j=3,4,
 \\\nonumber
 E^{w^{00}}_j(E^{w^{10}}_j)^{-1}&=&E^{w^{00}}_2(E^{w^{10}}_2)^{-1},\;\;j=3,4,5,
 \end{eqnarray}
supplemented by two of 
the remaining equations (\ref{Sec1:Eq_gen1_wj0}), for instance, those  
for $j=3,4$, $k=2$:
\begin{eqnarray}\label{D=2pick}
E^{w^{20}}_j(E^{w^{10}}_j)^{-1}&=&E^{w^{20}}_2(E^{w^{10}}_2)^{-1},
\end{eqnarray}
and by one (largely arbitrary) relation between the 8 matrix fields:
\begin{eqnarray}\label{D=2arbitraryF}
F(v,v_1,w_{00},w_{10},w_{20},w_{11},w_{12},w_{22})=0,
\end{eqnarray}
which is introduced in the same spirit as before.  
Other equations of the system (\ref{Sec1:Eq_gen1_v}) and 
(\ref{Sec1:Eq_gen1_wj0}) can be treated as a symmetries of eqs.
(\ref{D=2basic}, \ref{D=2pick}).

This procedure can be generalized to an arbitrary number $n$ of dimensions, introducing the 
following equation for $G$:
\begin{eqnarray}\label{Sec1:gen_g}
G_{x^j}=\sum_{k=1}^{n-3} \alpha_{jk} G_{x^k} ,\;\; j>n-3,
\end{eqnarray}
where $\alpha_{jk}$ are constant diagonal matrices and $n\ge 4$. The resulting 
nonlinear PDEs, possessing the same block structure as their lower dimensional analogies, 
will have dimensionality $n$ and, as we shall show in section \ref{Section:dimensionality}, will be characterized 
by a manifold of analytic solutions of dimension $n-2$.  

The possibility to increase the dimensionality of the PDEs and, at the same time, to increase 
proportionally the dimensionality of the manifold of solutions is due to the combined effect of the hypothesis:  
$\dim~{\mbox{ker}}\hat\Psi=1$ and of the introduction of the fields $w^{ij}$. Indeed, i) the property (\ref{Sec2:deg}) 
implies the multidimensional linear problems (\ref{Sec1:LU}) 
and, via equations (\ref{defA_D1},\ref{defA_D2}), 
the nontrivial mixing of the fields $v,v^1$ and $w^{ij}$; ii)  the matrix fields $w^{ij}$ are 
associated with the outer dressing function 
$G$, whose dimensionality can be increased without obstacles 
(see the ($n-3$)-dimensional equation (\ref{Sec1:gen_g}) ), while the matrix fields $v,v^1$ are  
associated with the matrix function $C$, an ingredient of the classical dressing method, 
whose dimensionality is severely constrained (see the 1-dimensional equations (\ref{Sec1:c_x})).
  
Note that, by construction, the derived systems possess higher symmetries:
eqs.(\ref{Sec1:Eq_short1},\ref{Sec1:Eq_short2}) with $j\ge 5$ and 
eqs.(\ref{D=2basic},\ref{D=2pick}) with $j\ge 6$.

We end this section elaborating on the dimensionality of the space of analytic solutions 
constructed in this section. 
Consider, as an illustrative example, the system of equations 
(\ref{Sec1:Eq_short3}),(\ref{rel_gamma}) in $n=4$ dimensions for the two matrix fields $w^{00},w^{10}$, and interpret $x^4$ 
as time variable. One can view the first equation $E^{w^{00}}_2(E^{w^{10}}_2)^{-1}=E^{w^{00}}_3(E^{w^{10}}_3)^{-1}$
as defining, in principle, $w^{10}$ in terms of $w^{00}$ and its partial derivatives $w^{00}_{x^j},~j=1,2,3$. 
Replacing this relation into the second equation $E^{w^{00}}_3(E^{w^{10}}_3)^{-1}=E^{w^{00}}_4(E^{w^{10}}_4)^{-1}$, 
one obtains a single equation for $w^{00}$, depending linearly on $w^{00}_{x^4}$ and nonlinearly on the other derivatives.    
Since, as we shall see in Sec.\ref{Section:dimensionality}, our dressing algorithm generates analytic solutions 
depending on an arbitrary 
matrix function of $n-2=2$ variables, the space of analytic solutions of (\ref{Sec1:Eq_short3}),(\ref{rel_gamma}) has 
dimension $n-2=2$. Similar arguments can be used for the systems (\ref{ex1}), (\ref{ex2}) and 
(\ref{D=2basic})-(\ref{D=2arbitraryF}), which exhibit a single equation with first order derivative with respect to 
$t=x^n$. Slightly different is a system (\ref{Sec1:Eq_short1})-(\ref{Sec1:condition_g}), in which two equations involve 
first order $x^4$-derivatives of the two functions $v$ and $w^{00}$. Accordingly, its solution space depends on {\it two} 
arbitrary functions of two variables, see (\ref{CPhi}) of \ref{Section:dimensionality}. Thus we have established 
 that the space of analytic solutions of all the examples of this section is ($n-2$)-dimensional.

\subsection{Compatibility of linear spectral problems versus nonlinear PDEs}

It is well-known that integrable PDEs arise as the compatibility of  
overdetermined systems of linear problems for some eigenfunction $U(\lambda;x)$. 
For instance, the N-wave system (\ref{Sec0:Nw1}) in 2+1 dimensions is the integrability 
condition for the Lax pair (\ref{Sec0:U_lin}). 

Such a picture is lost in our case, since there is no direct algebraic way to construct the nonlinear  
equations (\ref{Sec1:nl_j0}) as the compatibility condition of  the linear systems
(\ref{Sec1:LU}). Indeed, the compatibility between equations (\ref{Sec1:LU}) for $j\ne k$ leads to equation
\begin{equation}
\label{comp1}
\begin{array}{l}
U\big([B_k,L_{j1}v-[B_j,v_1]]-[B_j,L_{k1}v-[B_k,v_1]]\big)+(L_{21}U)\big(L_{k1}{A_j}-L_{j1}{A_k}\big)+ \\
(L_{21}U)_{x^1}\big(A_kB_j-A_jB_k\big)+(L_{21}U)_{x^k}A_j-(L_{21}U)_{x^j}A_k=0,
\end{array}
\end{equation}
from which one cannot infer anything, since the terms $(L_{21}U)_{x^j}$ and $(L_{21}U)_{x^k}$ are 
not independent matrix functions of $\lambda$, being expressible in terms of $L_{21}U$ and $U$.  
But, to obtain such expressions, one has to use additional structure, i.e., the dressing 
equation $\hat \Psi L_{j1}U=0$. Differentiating it with respect to $x^k$ and using equation (\ref{Sec1:x}), 
one obtains the homogeneous equation 
\begin{eqnarray}
\hat \Psi \big[(L_{j1}U)_{x^k} +U B_k \big(L_{j1}v-[B_j,v_1]\big)\big]=0,  
\end{eqnarray}
which, due to (\ref{Sec2:deg}), implies that 
\begin{eqnarray}\label{Hf}
(L_{j1}U)_{x^k} =-U B_k \big(L_{j1}v-[B_j,v_1]\big)+ (L_{21}U) {f^{jk}}, \;\;j\ne k,
\end{eqnarray}
where $f^{nk}$ are functions of $x$ only. Substituting these relations (with $j=2$) in (\ref{comp1}), one obtains:
\begin{equation}
\label{comp2}
\begin{array}{l}
U\big([B_k,E^v_j-E^v_2A_j]-[B_j,E^v_k-E^v_2A_k]\big)+\\
(L_{21}U)\big[L_{k1}{A_j}-L_{j1}{A_k}+f_{21}\big(A_kB_j-A_jB_k\big)+f_{2k}A_j-f_{2j}A_k\big]=0,
\end{array}
\end{equation}
and the independence of $U$ and $L_{21}U$ implies 
\begin{equation}
\label{comp3}
\begin{array}{l}
[B_k,E^v_j-E^v_2A_j]=[B_j,E^v_k-E^v_2A_k], \\
L_{k1}{A_j}-L_{j1}{A_k}+f_{21}\big(A_kB_j-A_jB_k\big)+f_{2k}A_j-f_{2j}A_k=0.
\end{array}
\end{equation}
We observe that equation (\ref{comp3}a) does not imply directly the wanted equations (\ref{Sec1:nl_j0}). 
To obtain them, one should consider, instead, the compatibility between the linear problem 
(\ref{Sec1:LU}) and equation (\ref{Hf}), which leads to the following equation:
\begin{eqnarray}\label{comp4}
UB_k\big(E^v_j-E^v_2A^j\big)+(L_{21}U)\big(A^j_{x^k}+f^{2k} A^j-f^{jk}\big)=0.
\end{eqnarray}
Again the independence of $U$ and $L_{21}U$ 
implies the relations $f^{jk}=f^{2k} A^j + A^j_{x^k}$, 
together with the wanted equations (\ref{Sec1:nl_j0}).   

Summarizing, the linear 3-dimensional problems (\ref{Sec1:LU}) contain only partial informations and their algebraic 
compatibility does not imply, alone, 
the nonlinear equations (\ref{Sec1:nl_j0}).

\section{Solution space}
\label{Section:Solution}
In the previous sections we have constructed partially integrable PDEs under the 
basic assumption that the integral equation (\ref{Sec1:U}) admits nontrivial homogeneous 
solutions. In this section we show i) how to choose the inner dressing functions in order to satisfy this 
assumption and, consequently, ii) how to construct the corresponding      
manifold of particular solutions of the partially integrable PDEs, expressed in terms of the 
dressing data $\Psi$, $\Phi$, $C$ and $G$. Note that, in our case, the novel 
dressing function $G$ appears, in comparison with the classical algorithm.

The solutions of equations (\ref{Sec1:Phi_x}) and (\ref{Sec1:c_x}) are:
\begin{eqnarray} \label{Sec2:Phi}
\Phi(\lambda;x)&=& \int \Phi_0(\lambda,k)e^{kB\cdot x} d k, \\
\label{Sec2:c}
C(\mu;x)&=&\int e^{q B\cdot x} C_0(q,\mu) d q,
\end{eqnarray}
where $k,q$ are scalar parameters and $\lambda,\mu$ are vector parameters of dimension $n-3$. Thus equations 
(\ref{Sec1:x}) yield:
\begin{eqnarray}\label{Sec0:Psi}
\Psi(\lambda,\mu;x)= 
\int \Phi_0(\lambda,k)e^{(k+q)B\cdot x}
 C_0(q,\mu)\frac{dk dq }{k+q}+\Sigma(\lambda,\mu),\;\;B_1=I. 
\end{eqnarray}
We remark that, in the case of $S$-integrable equations, the integration constant $\Sigma(\lambda,\mu)$ is chosen to be  
$\delta(\lambda-\mu)$. In our case, we need a special form for $\Sigma$ (see (\ref{def_A}) below). 

It is quite standard to assume that the measure $d\Omega(\lambda)$ have support on an open domain 
${\cal{D}}$ of the $\lambda$-space, and on a disjoint discrete set of points 
$D=\{b_1,\dots,b_M\}$, $D\cap {\cal{D}}=\emptyset$. Correspondingly, we use the following notation 
for the dressing functions.
\begin{eqnarray}\label{def_Phi_C}
\Phi(\lambda;x)&=&\left\{\begin{array}{ll}
\displaystyle \phi(\lambda;x)=\int \phi_0(\lambda,k) e^{k B\cdot x} dk, 
& \lambda \in {\cal{D}}, \cr
\displaystyle\phi_n(x)=\int \phi_{n0}(k) e^{k B\cdot x}dk, \;\;n=1,\dots,M, 
&\lambda \in D,
\end{array}\right.
\\\nonumber
\;\;\; 
C(\lambda;x)&=&\left\{\begin{array}{ll}
c(\lambda;x) = \int e^{q B\cdot x} c_0(q,\lambda) dq, 
& \lambda \in {\cal{D}}, \cr
c_n(x)=\int e^{q B\cdot x} c_{n0}(q) d q , 
&\lambda \in D,\;\;n=1,\dots,M, 
\end{array}\right.
\;\;\; 
\end{eqnarray}   
\begin{eqnarray}\label{def_G}
G(\lambda;x)&=&\left\{\begin{array}{ll}
g(\lambda;x), & \lambda\in {\cal{D}}, \cr
g_{n}(x), & \lambda \in {{D}},\;\; n=1,\dots,M  
\end{array}\right. 
\end{eqnarray}   
\begin{eqnarray}\label{def_U}
U(\lambda;x)&=&\left\{\begin{array}{ll}
u(\lambda;x), & \lambda\in {\cal{D}}, \cr
u_{n}(x), & \lambda \in {{D}},\;\; n=1,\dots,M,  
\end{array}\right. 
\end{eqnarray}   
and we choose $\Sigma(\lambda,\mu)$ in the form:
\begin{eqnarray}\label{def_A}
\Sigma(\lambda,\mu)&=&\left\{\begin{array}{ll}
\delta(\lambda-\mu), & \lambda, \mu\in {\cal{D}}, \cr
\sigma_{n}(\lambda), & \lambda \in {\cal{D}}
  \;\; \mu\in D,\;\;n=1,\dots,M \cr
\tilde \sigma_{n}(\mu), & \lambda\in D,
  \;\; \mu\in {\cal{D}},\;\;n=1,\dots,M \cr
\sigma_{nm}, &\lambda, \mu\in D, \;\;n,m=1,\dots,M. 
\end{array}\right. 
\end{eqnarray}   
Then equation (\ref{Sec1:U}) reduces to the following system of $M+1$ equations 
\begin{eqnarray}\label{Sec1:c_12}
\phi(\lambda;x)&=& 
\sum_{j=1}^M \int \phi_0(\lambda,k) e^{(k+q)B\cdot x} c_{j0}(q)\frac{dkdq}{k+q}  u_j(x)+ 
\\\nonumber
&&\int\limits_{{\cal{D}}} \phi_0(\lambda,k) e^{(k+q)B\cdot x} c_0(q,\mu) u(\mu;x)\frac{dkdqd\Omega(\mu)}{k+q} +
\sum_{j=1}^M \sigma_{j}(\lambda) u_j(x) +u(\lambda;x),\;\;\lambda\in{\cal D},\\\label{Sec22:Ub}
\phi_n(x)&=& 
\sum_{j=1}^M \int \phi_{n0}(k) e^{(k+q)B\cdot x} c_{j0}(q)\frac{dkdq}{k+q}  u_j(x)+ 
\\\nonumber
&&
\int\limits_{{\cal{D}}} \phi_{n0}(k) e^{(k+q)B\cdot x} c_0(q,\mu) u(\mu;x)\frac{dkdqd\Omega(\mu)}{k+q}   +
\\\nonumber
&&
\sum_{j=1}^M \sigma_{nj} u_j(x)+\sum_{j=1}^M \int\limits_{{\cal{D}}} \tilde \sigma_{n}(\mu) u(\mu;x)d\Omega(\mu),\;\;n=1,
\dots, M,
\end{eqnarray}
for the unknown matrix functions $u(\lambda;x),~\lambda\in{\cal D}$ and $u_j(x),~j=1,\dots,M$.

Once the solution is obtained, one constructs the matrix fields $v,~v^1,~w^{ij}$ using equations (\ref{Sec1:v},
\ref{Sec0:v1}, \ref{Sec0:w^ij}):
\begin{eqnarray}
\label{fields}
v(x)&=&\int\limits_{{\cal{D}}} c(\lambda;x) u(\lambda;x)
d\Omega(\lambda)+\sum\limits_{k=1}^M c_k(x) u_k(x),\\\nonumber
v^1(x)&=&\int\limits_{{\cal{D}}}\big[\partial_{x_1}
c(\lambda;x)\big] u(\lambda;x)
d\Omega(\lambda)+\sum\limits_{k=1}^M 
\big[\partial_{x_1} c_k(x)\big] u_k(x),\\\nonumber
w^{00}(x)&=&\int\limits_{{\cal{D}}} g(\lambda;x) u(\lambda;x)
d\Omega(\lambda)+\sum\limits_{k=1}^M g_k(x) u_k(x),\\\nonumber
w^{j0}(x)&=&\int\limits_{{\cal{D}}} \big[\partial_{x_j}
g(\lambda;x)\big] u(\lambda;x)
d\Omega(\lambda)+\sum\limits_{k=1}^M \big[\partial_{x_j}g_k(x)\big] 
u_k(x),\\\nonumber
w^{ij}(x)&=&\int\limits_{{\cal{D}}} \big[\partial_{x_i}\partial_{x_j}
g(\lambda;x)\big] u(\lambda;x)
d\Omega(\lambda)+\sum\limits_{k=1}^M 
\big[\partial_{x_i}\partial_{x_j}g_k(x)\big] 
u_k(x)
\end{eqnarray}

\subsection{The condition dim ${\mbox{ker}}\hat\Psi =1$.}
Now we have to provide the condition $\dim {\mbox{ker}}\hat\Psi =1$. 
We base our considerations on well known facts of the theory of linear integral operators. 
If the homogeneous equation 
\begin{eqnarray}
\label{HH}
\int\Psi(\lambda,\mu;x)H(\mu;x)d\Omega(\mu)=0
\end{eqnarray}
has a nontrivial solution, then its adjoint equation
\begin{eqnarray}
\label{tH}
\int\tilde H(\lambda;x)\Psi(\lambda,\mu;x)d\Omega(\lambda)=0
\end{eqnarray}
has a nontrivial solution as well. 
If $\dim{\mbox{ker}}\hat\Psi=1$, then the solution spaces of 
both equations (\ref{HH}) and (\ref{tH}) are one dimensional. 
  
In our case, equation (\ref{tH}) reads
\begin{eqnarray}
\int \tilde H(\lambda;x)\left[
\int^{x^1} d {x^1}' \Phi(\lambda;x') C(\mu;x') + \Sigma(\lambda,\mu)
\right] d\Omega(\lambda),\;\;x'=x|_{x^1\to {x^1}'}.
\end{eqnarray}
In view of the independence of $\Phi$ and $\Sigma$, this equation is splitted into two equations:
\begin{equation}\label{conditions}
\int \tilde H(\lambda;x) \Phi(\lambda;x')d\Omega(\lambda)=
\int \tilde H(\lambda;x) \Sigma(\lambda,\mu) d\Omega(\lambda)=0, 
\end{equation}
which have to be satisfied for all $x$, $x'$ and $\mu$. This means that $\tilde H$ is
independent of $x$ and, due to equation (\ref{Sec2:Phi}), the following two conditions 
must be satisfied:
\begin{eqnarray}
\label{condition:Phi}
\int \tilde H(\lambda) \Phi_0(\lambda,k)d\Omega(\lambda)=0,~~~\forall k,
\\\label{condition:G}
\int \tilde H(\lambda) \Sigma(\lambda,\mu) d\Omega(\lambda)=0,~~~\forall \mu 
\end{eqnarray} 
for the existence of a nontrivial solution of the homogeneous equation 
(\ref{HH}). 

It is important to remark that, at the same time, the condition (\ref{condition:Phi}) provides 
also the solvability of the inhomogeneous integral equation (\ref{Sec1:U}). Therefore no further constraint  
must be imposed.

We consider a particular way to 
satisfy conditions (\ref{condition:Phi}) and (\ref{condition:G}), choosing 
\begin{eqnarray}
\tilde H(\lambda)&=&\left\{\begin{array}{ll}
0, & \lambda \in {\cal{D}}, \cr
{\cal{A}}_j, & \lambda\in D,
\end{array}\right.
\end{eqnarray}   
where the matrices ${\cal{A}}_j$ are constant and nonsingular, so that the conditions (\ref{condition:Phi},\ref{condition:G}) 
are constraints only for the discrete parts of $\Phi$ and $\Sigma$:
\begin{eqnarray}\label{condition:A}
\sum_{j=1}^M {\cal{A}}_j \phi_j=
\sum_{j=1}^M {\cal{A}}_j \tilde \sigma_j(\mu)=
\sum_{j=1}^M {\cal{A}}_j \sigma_{jn}=0
,\;\;\mu\in{\cal D},\;\;n=1,\dots,M.
\end{eqnarray}

Due to (\ref{condition:A}), we have only  $(M-1)$  independent equations in the system 
(\ref{Sec22:Ub}) and, consequently, the solutions $u_j(x)$ are constructed up to an arbitrary function $f(x)$. 

We remark that terms containing $\sigma_j$, $\sigma_{ij}$ and $\tilde \sigma_j$ may disregarded in equations 
(\ref{Sec1:c_12},\ref{Sec22:Ub}). Indeed, 
 the terms containing $\sigma_j$ in equation (\ref{Sec1:c_12})
 can be incorporated in the first term of the RHS. Similarly, the  
 terms with $\sigma_{ij}$ and $\tilde \sigma_j$ in the equation (\ref{Sec22:Ub})
can be incorporated in the first and second terms of this equation. Thus we set $\sigma_j=\tilde \sigma_j=\sigma_{ij}=0$ 
without loss of generality. 
\subsection{Degenerate kernel.}

The system of linear equations (\ref{Sec1:c_12},\ref{Sec22:Ub}), supplemented by the conditions (\ref{condition:A}), 
has a rich manifold of solutions. To construct explicit solutions, we choose, as 
usual, a degenerate kernel:
\begin{eqnarray}
c_0(q,\mu) = \sum_{j=1}^{\tilde M} 
\tilde c_{1j}(q) \tilde c_{2j}(\mu). 
\end{eqnarray} 
In this case, equations (\ref{Sec1:c_12},\ref{Sec22:Ub}) reduce to
the following linear system of $M+\tilde M$ equations 
\begin{eqnarray}\label{Sec1:c_12_red}
\tilde \phi_n(x)&=& 
\sum_{j=1}^M\nu_{nj}(x)u_j(x)+\sum_{j=1}^{\tilde M}\tilde \nu_{nj}(x) \tilde u_j(x)+\tilde u_n(x),~~n=1,\dots,\tilde M,
\\\label{Sec22:Ub_red}
\phi_n(x)&=&\sum_{j=1}^M\rho_{nj}(x)u_j(x)+\sum_{j=1}^{\tilde M}\tilde\rho_{nj}(x)\tilde u_j(x),~~~~~~~~~~n=1,\dots, M,
\end{eqnarray}
for the matrix fields $u_j(x),~\tilde u_k(x),~j=1,\dots,M,~k=1,\dots,\tilde M$, 
where:
\begin{eqnarray}
\tilde u_k(x) = \int \tilde c_{2k}(\lambda) u(\lambda;x)d\Omega(\lambda),\;\;
\end{eqnarray}
and where the given coefficients $\nu_{nj},\tilde \nu_{nj},\rho_{nj},\tilde\rho_{nj},\tilde \phi_n$ are defined in terms 
of the spectral functions:
\begin{equation}
\label{def_rho}
\ba{l}
\nu_{nj}(x)=\int \tilde\phi_{n0}(k) e^{(k+q)B\cdot x}c_{j0}(q)\frac{dkdq}{k+q},~~
\tilde \nu_{nj}(x)=\int\tilde\phi_{n0}(k) e^{(k+q)B\cdot x}\tilde c_{1j}(q)\frac{dkdq}{k+q}, \\
~~ \\
\rho_{nj}(x)=\int \phi_{n0}(k) e^{(k+q)B\cdot x} c_{j0}(q)\frac{dkdq}{k+q},~~
\tilde\rho_{nj}(x)=\int \phi_{n0}(k) e^{(k+q)B\cdot x}\tilde c_{1j}(q)\frac{dkdq}{k+q}, \\
~~ \\
\tilde \phi_n(x) = \int \tilde c_{2n}(\lambda)\phi(\lambda;x)d\Omega(\lambda),\;\; \tilde\phi_{n0}(k)=
\int\limits_{{\cal{D}}} \tilde c_{2n}(\lambda)\phi_0(\lambda,k) d\Omega(\lambda).
\ea
\end{equation}
This algebraic system is obtained, as usual, applying the operator 
$\int\limits_{{\cal{D}}} c_{2n}(\mu;x)d\Omega(\mu) \cdot$ 
to (\ref{Sec1:c_12}). 

Having constructed, from (\ref{Sec1:c_12_red}), (\ref{Sec22:Ub_red}), the $u_s(x)$ and the $\tilde u_s(x)$, 
one obtains the eigenfunction $u(\lambda;x)$ via the formula:
\begin{eqnarray}
u(\lambda;x)=\int \phi_0(\lambda,k)e^{k B\cdot x} dk -\sum_{j=1}^M \rho_j(\lambda;x)u_j(x)
-\sum_{j=1}^{\tilde M} \tilde\rho_j(\lambda;x)\tilde u_j(x),
\end{eqnarray}
where:
\begin{eqnarray} 
\rho_j(\lambda;x)=\int \phi_0(\lambda,k)e^{(k+q) B\cdot x} c_{j0}(q)\frac{dk dq}{k+q}, \\
\tilde\rho_j(\lambda;x)=\int \phi_0(\lambda,k)e^{(k+q) B\cdot x} \tilde c_{1j}(q)\frac{dk dq}{k+q}.
\end{eqnarray}
At last, one constructs the matrix fields $v,v^1,w^{ij}$, solutions of the nonlinear PDEs of Sec. 3.2, from 
equations (\ref{fields}).

\subsection{Fixing the arbitrary function $f(x)$}
\label{Section:Fixing}

Due to the constraint (\ref{condition:A}), the solutions $u_j$ and $\tilde u_j$ of the algebraic system (\ref{Sec1:c_12_red}) 
depend linearly on an arbitrary matrix function $f(x)$. Then, via (\ref{fields}), also the fields $v,v^1,w^{ij}$ 
depend linearly on this arbitrary function. Such an arbitrary function is completely fixed by the largely 
arbitrary relation among the fields.  

To be more concrete, let us illustrate all these facts in the simplest case:   
$c_0(q,\mu)=0$, $M=2$. The constraint (\ref{condition:A}) implies that 
$\rho_{2j}=-{\cal A}^{-1}_2{\cal A}_1\rho_{1j},~j=1,2$; then, from  
the homogeneous version of (\ref{Sec22:Ub_red}):
\beq
\ba{l}
0=\rho_{11}(x)H_1(x)+\rho_{12}(x)H_2(x), \\
0=\rho_{21}(x)H_1(x)+\rho_{22}(x)H_2(x),
\ea
\eeq 
one verifies that the second equation is consequence of the first, 
while the first equation admits the solution
\beq
H_1(x)=\rho^{-1}_{11}(x)f(x),\;\;\;H_2(x)=-\rho^{-1}_{12}(x)f(x)
\eeq
depending linearly on the arbitrary matrix function $f(x)$ (compare with (\ref{Sec1:Unhom})). The general solution of the 
inhomogeneous algebraic system (\ref{Sec22:Ub_red}) is then given by:
\beq
\label{explicit}
\ba{l}
u_1(x)=\frac{1}{2}\rho^{-1}_{11}(x)\phi_1(x)+\rho^{-1}_{11}(x)f(x), \\
~~ \\
u_2(x)=\frac{1}{2}\rho^{-1}_{12}(x)\phi_1(x)-\rho^{-1}_{12}(x)f(x), 
\ea
\eeq
depending on the arbitrary matrix function $f(x)$ in a linear way as well (compare with (\ref{Sec1:Uninhom})). Consequently, 
such a linear dependence on $f(x)$ will appear, via (\ref{fields}), also in the matrix fields $v,v^1,w^{ij}$. 

We remark that one could always identify $f(x)$ with one of the $u_j$'s, say, with $u_1(x)$, obtaining 
\beq
u_1(x)=f(x),~~~~u_2(x)=\rho^{-1}_{12}(x)\left[\phi_1(x)-\rho_{11}(x)f(x)\right];
\eeq
this identification, which clearly leads to a less symmetric formula than (\ref{explicit}), seems to become more 
convenient when $M>2$. 

Now we show, always in the simplest case: $c_0(q,\mu)=0$, $M=2$, how the arbitrary function $f(x)$ gets 
fixed imposing the relation  
(\ref{Sec1:condition00}), which we choose in one of the forms pointed after equation (\ref{Sec1:fields_f}).

\vskip 5pt
\noindent
{\bf Equations of Sec.\ref{Section:simplest}}. For the equations of this section, choose $g(\lambda)=\delta(\lambda-a)$. 

Then, using equation (\ref{fields}) for $w^{00}$, 
the constraint (\ref{Sec1:condition0}) becomes the linear equation
\beq
\ba{l}
h_0(x)+
\left[h_1(x)\rho^{-1}_{11}(x)+h_2(x)\rho^{-1}_{12}(x))\right]\frac{\phi_1(x)}{2}+ \\
\left[h_1(x)\rho^{-1}_{11}(x)+h_2(x)\rho^{-1}_{12}(x))\right]f(x)=\exp\left(\sum_{i=1}^{n} {a_i x^i}\right)
\ea
\eeq
for $f(x)$, where
\beq
h_j(x)=-\int\limits_{\cal D}g(\lambda;x)\rho_j(\lambda;x)d\Omega(\lambda)+
g_j(x),~~j=1,2,\;\;h_0(x)=\int\limits_{\cal D}g(\lambda;x)\phi(\lambda;x)d\Omega(\lambda)
\eeq 
yielding the following explicit formula for $f$:
\beq
\ba{l}
f(x)=\left[h_1(x)\rho^{-1}_{11}(x)+h_2(x)\rho^{-1}_{12}(x))\right]^{-1}
\Big[\exp\left(\sum_{i=1}^{n} {a_i x^i}\right)- \\
h_0(x)-
\left[h_1(x)\rho^{-1}_{11}(x)+h_2(x)\rho^{-1}_{12}(x))\right]\frac{\phi_1(x)}{2}\Big], 
\ea
\eeq
Analogously, using equation (\ref{fields}) for $v^1$, the constraint (\ref{Sec1:condition01}) 
becomes a linear equation for $f(x)$, whose explicit solution is  
\beq
f(x)=\left[h_1(x)\rho^{-1}_{11}(x)+h_2(x)\rho^{-1}_{12}(x))\right]^{-1}
\left[\gamma(x)-\left[h_1(x)\rho^{-1}_{11}(x)+h_2(x)\rho^{-1}_{12}(x))\right]\frac{\phi_1(x)}{2}\right],
\eeq
where now:
\beq
h_j(x)=\int q e^{q B\cdot x} c_{j0}(q) d q,~~~j=1,2.
\eeq
 
\vskip 5pt
\noindent
{\bf Equations of Sec.\ref{Section:general}}. In this case, equation (\ref{Sec1:gen_g}) implies that 
 \begin{eqnarray}\label{Sec:Dr:gg}
G(\lambda;x)=\int\limits_{\cal D}\exp\left[\sum_{j=1}^{n-3}
\lambda'_j \left(x^j+\sum\limits_{k=n-2}^{n}\alpha_{kj}x^k\right)\right]G_0(\lambda',\lambda)d\Omega(\lambda').
\end{eqnarray} 

If, in particular, $n=4$, as in equation (\ref{g-x}), 
then the reduction $G=C$ is admissible, identifying
$\alpha_j=\alpha_{j1}=B_j$ and $G_0=C_0$ (see (\ref{Sec2:c})), 
and one obtains the nonlinear PDEs (\ref{Sec1:Eq_short3}), 
supplemented by the relation (\ref{D=2arbitraryF}). 

If this relation is an arbitrary linear 
relation between fields, it leads to a linear equation for $f(x)$. For instance, if we choose (\ref{rel_gamma}),   
then, using the equation (\ref{fields}) for $w^{11}$, the constraint  
becomes a linear equation for $f(x)$, whose explicit solution is  
\beq
\ba{l}
f(x)=\left[h_1(x)\rho^{-1}_{11}(x)+h_2(x)\rho^{-1}_{12}(x))\right]^{-1}
\Big[\gamma(x)- \\
\int\limits_{\cal D}g_{x^1x^1}(\lambda;x)\phi(\lambda;x)d\Omega(\lambda)-
\left[h_1(x)\rho^{-1}_{11}(x)+h_2(x)\rho^{-1}_{12}(x))\right]\frac{\phi_1(x)}{2}\Big], 
\ea
\eeq
where:
\beq
h_j(x)=-\int\limits_{\cal D}g_{x^1x^1}(\lambda;x)\rho_j(\lambda;x)d\Omega(\lambda)+{g_j}_{x^1x^1}(x),~~j=1,2.
\eeq
In a similar way, one can treat more general relations in higher dimensions $n$.
\subsection{Dimensionality of the solution space}
\label{Section:dimensionality}
The dimensionality of the space of analytic solutions generated by our dressing scheme is essentially defined 
by the dimensionality of two expressions 
\beq
\label{CPhi}
\ba{l}
\int C(\lambda;x)\Phi(\lambda;x)d\Omega(\lambda),\;\;\;\; 
\int G(\lambda;x)\Phi(\lambda;x)d\Omega(\lambda).
\ea
\eeq 
The first term, involving the ``inner'' dressing functions $C(\lambda;x)$ and 
$\Phi(\lambda;x)$, appears also in the classical dressing;  
its dependence on the space-time coordinates is severely constrained and, consequently, it carries dimensionality 2 
(see (\ref{dim2}) and the considerations made there). The second term involves the ``outer'' dressing function 
$G(\lambda;x)$, a novel feature of our dressing 
procedure; its dependence on the space-time coordinates is instead largely arbitrary, playing a crucial role 
in increasing the dimensionality of the solution space  through the following novel 
mechanism. 

As we have seen in Sec.\ref{Section:Fixing}, from the largely arbitrary relation among the fields one construct $f(x)$ in 
terms of the spectral representations of the fields $v,v^1,w^{ij}$. If such relation involves the fields $w^{ij}$, 
whose spectral representations involve expressions like (\ref{CPhi}b), then 
the dimensionality of $f(x)$ is not severely constrained. Since the matrix fields $v,v^1,w^{ij}$, solutions of 
our PDEs, depend linearly on $f(x)$, their analytic solution space is not severely constrained too. 
Using this argument, it is possible to establish easily the dimensionality of the 
space of analytic solutions of our PDEs.    

In Sec.\ref{Section:simplest}, the first term has higher dimensionality than the second, since
$G$ does not depend on $x$; it follows that the dimensionality of the space of analytic solutions 
of the 4 dimensional PDEs constructed there is 2, like for integrable PDEs in 2+1 dimensions. 
 
In Sec.\ref{Section:general}, the outer dressing function is (\ref{Sec:Dr:gg}) and the second term in 
equation (\ref{CPhi}) reads (choosing $G_0(\lambda',\lambda)=\delta(\lambda'-\lambda)$):
\begin{eqnarray}
\int\limits_{\cal D}\exp{\left[\sum_{j=1}^{n-3}
\lambda_j \left(x^j+\sum\limits_{k=n-2}^{n}\alpha_{kj}x^k\right)\right]}
\Phi_0(\lambda,k)\exp{\left[k\left(B\cdot x\right)\right]}dkd\Omega(\lambda).
\end{eqnarray}
Since $\lambda$ is a vector parameter of dimension $n-3$, and $k$ is a scalar parameter, the above 
expression has dimension $n-2$, being parameterized by the arbitrary function $\Phi_0(\lambda,k)$ 
of $n-2$ variables. Consequently, $n-2$ is the dimension of the constructed $f(x)$, and of all the 
fields appearing in the nonlinear PDEs. 

Summarizing, the solutions we constructed depend on an arbitrary function of $n-2$ variables (\ref{CPhi}b) and 
on an arbitrary function of $2$ variables  (\ref{CPhi}a). Then, in the exceptional case $n=4$, the solutions 
depend on $2$ arbitrary functions of $2$ variables. This conclusion is valid for all the examples presented in the  paper.

\section{Conclusions}
\label{Section:conclusion}
In this paper we have generalized the dressing method to construct systems of nonlinear PDEs 
in $n$ dimensions ($n>3$) i) possessing a manifold of analytic solutions of dimension $n-2$ (a very large, 
but not complete, manifold), and ii) possessing 
higher symmetries. But the constructed PDEs do not seem to be the compatibility condition for overdetermined systems of linear 
PDEs, a characterizing feature of completely integrable systems in lower dimensions. 

The above properties indicate that they 
are examples of partially integrable PDEs in multi-dimension possessing a very large, but not complete, space of 
analytic solutions.  

A natural generalization of the algorithm presented in this paper consists in studying the case in which the integral operator 
$\hat\Psi$ of the dressing problem exhibits a higher dimensional kernel:
 \begin{eqnarray}
 \dim {\mbox{ker}} \hat\Psi =D^{ker} >1.
 \end{eqnarray}  
In this case, equation (\ref{Sec1:LU}) is replaced by  
\begin{eqnarray}
L_{m1}U(\lambda;x)+\sum_{n=2}^{D^{ker}+1} (L_{n1} U(\lambda;x)) A^{mn}(x) =0, \;\;\;\;m>D^{ker}+1,
\end{eqnarray}
and one needs $D^{ker}$ conditions on $U$ to define the functions 
 $A^{mn}$. The study of the structure of the associated partially integrable equations, and of the 
dimensionality of the associated analytic solution space is postponed to future investigations. 

\vskip 10pt

The authors thank the referees for useful comments.
 This work was supported by the INTAS Young Scientists Fellowship Nr. 04-83-2983,
 Grants RFBR 04-01-00508 and NSh 1716-2003.

\end{document}